\documentclass[%
 amsmath,amssymb,
 aps,
]{revtex4-2}

\usepackage{graphicx}
\usepackage{dcolumn}
\usepackage{bm}

\usepackage{hyperref}

\usepackage{amsmath}
\usepackage{amssymb}
\usepackage{epstopdf, epsfig}

\usepackage{color, colortbl}

\usepackage[utf8]{inputenc}
\usepackage[T1]{fontenc}
\usepackage{mathptmx}

\newcommand{\blue}[1]{\color{black}{#1}\color{black}{}}

\usepackage{soul}
\usepackage[normalem]{ulem}
\usepackage{xcolor}

\newif\ifshowcomments

\showcommentstrue

\ifshowcomments
\newcommand{\KSH}[1]{{\color{red}{#1}}} 
\newcommand\KSHo{\bgroup\markoverwith{\textcolor{red}{\rule[0.5ex]{2pt}{0.4pt}}}\ULon}
\else
\newcommand{\KSH}[1]{}
\newcommand\KSHo[1]{}%
\fi

\begin{document}

\preprint{APS}

\title{Mode Competition in a Plunging Foil with an Active Flap: A Multi-Scale Modal Analysis Approach}

\author{Tso-Kang Wang}
\author{Kourosh Shoele}%
 \email{kshoele@eng.famu.fsu.edu}
\affiliation{%
 Department of Mechanical Engineering, Joint College of Engineering Florida State University-Florida A\&M University, Tallahassee, Florida 32310, USA
}%


\begin{abstract}
The flow-induced flutter has a significant role in aircraft stability, renewable energy extraction, animal locomotion, among many other applications. While being a ubiquitous phenomenon, the control of the flutter response has been primarily limited to simplified systems and, often, with the help of linear inviscid flow theories. In this paper, we numerically investigate how the plunging response of a foil can be regulated using an active flap to improve structural safety or enhance the energy extraction efficiency of the foil with a tightly coupled fluid-structure interaction algorithm. A broad range of foil and flap settings was tested and their flow dynamics have been investigated. A novel multi-scale modal analysis technique suitable for fluid-structure interaction system successfully isolates the active flap-induced and the flow-induced modes. It is observed that the competition between these two modes dictates the plunging response  of the foil. The active flap can modulate the leading edge vortex shedding with larger flapping amplitude and regulate the foil heaving motion. The ratio of the competing modal energy is proposed to evaluate the control efficacy of the morphing surface, and the onset of the lock-in is associated with the ratio approaching unity. It is shown that the morphing flap is a good candidate for active flow control.
\end{abstract}

\maketitle


\section{Introduction}
Flow-induced flutter has been the subject of continual fluid dynamic research since the failed attempt to fly a monoplane by Samuel P. Langley in 1903. In his text, Bisplinghoff \cite{bisplinghoff2013aeroelasticity} defined the aeroelasticity as {\it "the phenomenon which exhibits appreciable reciprocal interactions (static or dynamic) between aerodynamic forces and the deformations induced in the structure"}. 
The physics of flow-induced flutter is composed of interactions between the flow and elastic or inertial structures\cite{collar1946expanding}. 
Two primary mechanisms determine the fluttering status of a foil: the coupling of the torsional and heaving modes of the airfoil; and the flow separation and reattachment on the foil surface \cite{hansen2007aeroelastic}. Under proper conditions, the fluttering response can lead to self-sustained limit cycle oscillations and even induce instabilities \cite{ducoin2013hydroelastic, poirel2011computational, zhu2020nonlinear}. In the case of instability, the fluttering motion caused by the aeroelastic interaction presents a severe threat to the airfoil structural integrity. With the development of faster and lighter airplanes, aeroelastic instability has become even a more critical design factor \cite{kehoe1995historical, mukhopadhyay2003historical}. Structures other than aerial systems can also be affected by aeroelasticity. A famous instance is the collapse of Tacoma Narrows Bridge due to stall flutter \cite{amman1941failure}.

Flow-induced elastic oscillation is also a critical physical mechanism in many biological systems. Oscillating foils are often applied as a prototype model of animal locomotion. For example, the beating of the caudal fins of thunniform fishes such as tuna have been studied using pitching and heaving foil configurations \cite{fierstine1968studies, sfakiotakis1999review, zhu2008propulsion}. The hovering motion of insects or miniature flying systems can be mathematically described with rapidly oscillating foil-shaped wings. This model has been adopted to study the efficiency of the flapping foils \cite{freymuth1990thrust, liu1998numerical, read2003forces, sum2019scaling, wu2020review}. Furthermore, in the past decade, the potentials of utilizing the flow-induced elastic response for energy harvesting have been explored for various conditions \cite{zhu2009mode,abdelkefi2016aeroelastic, rostami2017renewable}. Energy-generating devices, such as a piezoelectric component attached to a foil, can harvest energy from the flow through the vibration. Flapping foils as energy harvesters are environmentally friendly with fewer maintenance requirements and can be easily integrated into existing vehicle designs \cite{xiao2014review, li2016energy}. 

Different flow control techniques have been proposed to suppress the vibration or regularize the airfoil's dynamic stall response. Passive control methods have been suggested to suppress vibration. As an example from many other studies, Lee et al. \cite{lee2007suppression} explored whether the redistribution of the mechanical energy with nonlinear energy sinks could postpone the cascade of heaving and pitching modes of the airfoil, and Fatimah et al. \cite{fatimah2003suppressing} demonstrated that the vibration of the structure could be alleviated by adjusting the stiffness between the oscillating object and an absorbing mass. Recently smart materials have been used to produce continuous surface deflection of an airfoil and modify aerodynamic characteristics of the system \cite{giurgiutiu2000review, barbarino2011review, moored2005analysis}. Among different flow control techniques, conceivably, the most feasible engineering solution is still an active flap at the leading edge or trailing edge. Block et al. \cite{block1998applied} used Theodorsen's flow Theorem \cite{theodorsen1935general} to identify a controlling action that can stabilize the system by adjusting the angle of the trailing edge flap. Wang et al. \cite{wang2011model} further used a multi-input system with active control surfaces at both leading and trailing edges and designed a full-state feedforward/feedback controller with a high-gain observer. Experimentally, Platanitis et al. \cite{platanitis2004control} employed a geometrically nonlinear controller designed via Lie algebraic methods on a dual-control surface airfoil to suppress the oscillation under optimal conditions. Medina et al.\cite{medina2020separated} experimentally investigated the physics of flow separation at the leading edges due to rapid trailing-edge flap deflection. 

Many fundamental questions remain unsettled despite previous research on using control surfaces to modulate flow-induced flutter. The model-based control methods developed for active control surfaces are often centered around inviscid flow models such as Theodorsen's theory, in which a foil is represented with a two-dimensional inviscid plate \cite{ohta1989active, tang1998limit, zhang2016continuous}. These control algorithms consider the fluid-structure interaction (FSI) dynamics as a change in the flow momentum due to small amplitude motion around the structure's mean position. Therefore, they cannot capture the flow change associated with the large structural deformation, unsteady vortex separations and nonlinear vortex-body interaction. Furthermore, few reduced-order modeling techniques are available that simultaneously consider both structure and fluid responses. For example, Goza and Colonius \cite{goza2018modal} combined fluidic and structural energy to construct hybrid proper orthogonal decomposition (POD) modes to correlate the structural motion and the induced flow features. Liberge et al. \cite{liberge2010reduced} interpolate the time-variant grid to a fixed grid to perform POD on a global velocity field. Menon and Mittal \cite{menon2020dynamic} fix the moving frame and correct the dynamic mode decomposition (DMD) modes with the frequency spectrum of the fixed shape foil motion. \blue{Still, certain limitations apply to these methods, like the solid and fluid modes are not correlated or the requirement of the knowledge about the periodic structural behavior. Here, we proposed a more general method to inspect the modal content of the flow-induced fluttering response of morphing bodies. A conformal mapping technique is used to generate unique energy-based geometrical weights, which enable the modal analysis to be employed for the FSI problems with morphing bodies \cite{wang2021geometrically}. The method has no assumption on the geometry and can be applied posterior to the data.}

This paper employs the proposed FSI algorithm based on conformal geometry mapping to investigate how an active morphing flap modifies the flow field around a foil. In addition, a reduced-order model of the flap and foil is developed, which, unlike many other reduced order methods, can encapsulate the deforming geometry of the bodies, the nonlinear fluid force feedback, and the flow field created by the motion of the foil. This offers a more refined understanding of the flow-induced fluttering phenomena and can be employed as a fast surrogate model for future controller designs. We also introduced a multi-scale modal analysis technique \cite{mendez2019multi} along with the conformal geometrical weighting \cite{wang2021geometrically} to isolate the competing modes in the FSI system. The main contribution of this work can be summarized as (1) developing a high-fidelity body-conformed FSI algorithm to investigate foil and flap system; (2) studying the flow-induced plunging response of a foil with an actively oscillating morphing flap over a wide range of structural parameters; (3) extending the multi-scale proper orthogonal decomposition (mPOD) to FSI system wherein the geometrical weighting is used to explain how the mode competition between the flap-induced and flow-induced modes dictates the fluttering motion of the foil.

The rest of the paper is organized as follows: \S\ref{sec:methodology} discusses the fluid-structure interaction method to model a foil with an active flap system and the geometrically weighted multi-scale POD method. In \S\ref{sec:qualitative} we present the results of the system following by a discussion about the flap-induced and flow-induced flow features in \S\ref{sec:competition}. At the end, in \S\ref{sec:conclusion}, we conclude the current work and present potential future directions.

\section{\label{sec:methodology}Methodology}
\begin{figure}
	\centering
	\includegraphics 
	{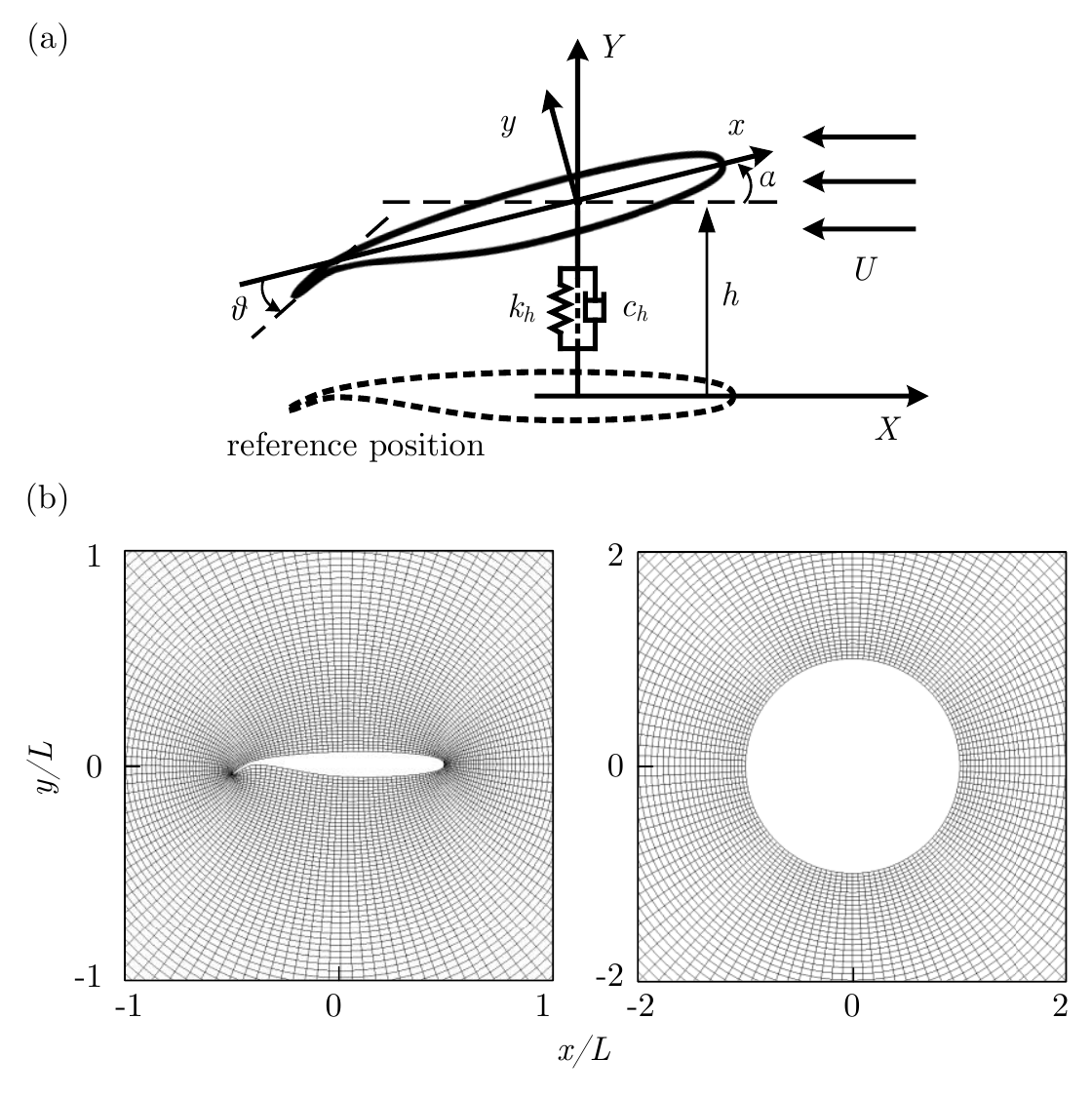}
	\caption{(a) The cross-section of the foil and the model parameters. (b) Sample computational grid around the foil-flap model and the grid mapping to a unit circle.}
	\label{fig:1}
\end{figure}

This section will introduce the numerical model and the modal analysis method adopted to investigate the aeroelastic response of an airfoil with an active flap. A two-dimensional EET airfoil immersed in uniform ambient flow is chosen as a model structure. The EET high-lift airfoil\cite{van2002aerodynamic} was developed by NASA and has been through extensive experimental tests \cite{morgan2002experimental} to understand how the multi-sectional foil design affects the lift generation. It has also been employed for studying the controllable leading and trailing edges to improve the performance of the airfoil using linear models \cite{nissim1971flutter,reddy2007multi}. Here, we extend these works by incorporating the continuous shape changes of the morphing foil and its flow-induced plunging motions. As shown in Fig. \ref{fig:1}(a), the flow-induced fluttering characteristics of the foil are represented with a translational spring-damper system. We focus on the heaving motion of the foil and assume it has a fixed pitching angle. The heaving displacement of the virtual attachment point of the spring system is $h$, the geometrical angle of attack (AoA) of the foil relative to the incoming flow is $\alpha$ , and the angle between the flap and the mean chord line of the foil is $\theta$. Comparable models have been used previously to study the flow-induced fluttering response \cite{guglielmini2004propulsive, zhu2009mode, chae2013dynamic, ducoin2013hydroelastic}. In the current study, the flap angle $\theta$ is actively adjusted to represent the morphing flap as the simplest mode of the geometrical controller.

\subsection{\label{subsec:numeric} Numerical Method} 
The numerical method is composed of three main parts: setup of the computational mesh, calculation of the dynamic forces, and the fluid-structure interaction procedure, which is explained in detail and verified in a prior publication \cite{wang2021geometrically}. Thereby, We only briefly outline the key components in the following.

A body-fitted grid with a higher resolution near the body is employed to model the highly nonlinear flow around the foil. Here, a conformal geometrical mapping technique \cite{pozrikidis2011introduction, eldredge2019mathematical} is used to generate a time-variant body-fitted grid. The body contour is segmented in the foil attached coordinate system into points $z_j=x_j+i y_j$. The body exterior is then mapped to the exterior of a circle with the radius $\lambda$. The cylindrical coordinate along the circle is defined as $\zeta=\lambda e^{i\phi}$ with $\phi$ denoting the azimuth angle. The mapping between the two coordinate systems is described with the Laurent series expansion, 
\begin{equation}
z(\phi)=\lambda e^{i\phi}+a_0+\displaystyle\sum_{n=1}^{\infty} a_ne^{-i n\phi},
\end{equation}
where the coefficients $a_n$ contain the radial terms $a_n\equiv \lambda^n b_n$ with $b_n$ being constant coefficients. 
We can then obtain the mapping coefficients and optimal coordinates by employing the orthogonality properties of series terms and minimizing the error between the mapped and actual contour points. The grid is constructed in the transformed domain then mapped to the physical domain with the same set of coefficients acquired from the body contour.

A sample grid generated for the foil with this algorithm is illustrated in Fig. \ref{fig:1}(b). The grid is stretched in radial direction logarithmically to increase its resolution near the surface where large velocity gradients are anticipated. The cylindrical grid in $\zeta$ plane is shown in Fig. \ref{fig:1}(b). During the simulation, the mesh is updated based on the instantaneous flap angle. The conformal mapping technique also offers a time-invariant grid in the transformed space, allowing the modal analysis techniques. This will be discussed later in \S\ref{subsec:mPOD}.

The flow field is modeled with a modified Navier--Stokes equation based on the vorticity-stream function $(\omega-\psi)$ formulation in the transformed domain. An implicit, second-order finite difference approach proposed by Guglielmini et al.\cite{guglielmini2004propulsive} and Zhu et al.\cite{zhu2009mode} is employed to solve the two-dimensional Navier-Stokes equation,
\begin{equation}
\frac{\partial\omega}{\partial t}+\frac{1}{\sqrt{J}}\left[v_r\frac{\partial\omega}{\partial r}+\frac{v_\phi}{r}\frac{\partial\omega}{\partial\phi}\right]\,=\,\frac{1}{Re \, J}\left[\frac{\partial^2\omega}{\partial r^2}+\frac{1}{r} \frac{\partial \omega}{\partial r}+\frac{1}{r^2}\frac{\partial^2\omega}{\partial\phi^2}\right]
\end{equation}
\begin{equation}
\frac{\partial^2\psi}{\partial r^2}+\frac{1}{r}\frac{\partial\psi}{\partial r}+\frac{1}{r^2}\frac{\partial^2\psi}{\partial\phi^2}\,=\,-J\omega,
\end{equation}	
where $Re$ is the Reynolds number and $J$ is the temporally and spatially varying determinant of the Jacobian of $\zeta=\xi+i\eta \Rightarrow z=x+iy$ transformation. The radial and azimuthal flow velocities, $v_r$ and $v_\phi$ respectively, include the time-dependent effect of the transformation. The no-slip boundary condition is applied on the foil surface and uniform flow condition with $\vec{v}=[U_\infty,0]^T$ and $\omega=0$ is applied at the far-field boundaries.

After the vorticity and stream function are solved in the transformed domain, the inverse mapping transformation is used to calculate flow data in the physical domain. The non-dimensionalized pressure, hydrodynamic force, and hydrodynamic moment exerted on the foil are then calculated using the following equations, integrated along the foil boundary in the transformed domain:
\begin{equation}
	p(\phi)=p_0+\int_{0}^{\phi}\left(\frac{\partial p}{\partial X}\frac{\partial X}{\partial \phi}+\frac{\partial p}{\partial Y}\frac{\partial Y}{\partial \phi}\right)d\phi,
	\end{equation}
	\begin{eqnarray}
	\frac{F_X}{\rho U_{\infty}^2 L}&=&\int_{0}^{2\pi}-p\frac{\partial Y}{\partial \phi}+\frac{2}{Re}\left[\frac{\partial Y}{\partial \phi}\frac{\partial U}{\partial X}-\frac{1}{2}\frac{\partial X}{\partial \phi}\left(\frac{\partial U}{\partial Y}+\frac{\partial V}{\partial X}\right)\right]d\phi,\\
	\frac{F_Y}{\rho U_{\infty}^2 L}&=&\int_{0}^{2\pi}p\frac{\partial X}{\partial \phi}+\frac{2}{Re}\left[\frac{1}{2}\frac{\partial Y}{\partial \phi}\left(\frac{\partial U}{\partial Y}+\frac{\partial V}{\partial X}\right)-\frac{\partial X}{\partial \phi}\frac{\partial V}{\partial Y}\right]d\phi,
\end{eqnarray}
where $U$ and $V$ are the non-dimensional flow velocity components in the $X$ and $Y$ directions along the body surface. 

The structural heaving velocity, $\dot{h}_n$, is updated, through a tightly coupled algorithm, using the calculated forces and moments:
\begin{equation}
	m_f\,\ddot{h}\,+\,c_h\,\dot{h}\,+\,k_h\,h\,=\,F_Y ,
\end{equation}
where $m_f$ is the mass of the foil, $c_{h}$ and $k_{h}$ are the damping and stiffness coefficients for heaving motions, respectively. \blue{The vibration caused by the change of center of the mass when the flap moves is ignored in current study as its contribution is negligible with small flap mass and movement. For applications with larger deformation or when the flap occupies significant body weight, however, this should be taken into account.}

The fluid equations are discretized in $r$, $\phi$ and $t$ and solved in the transformed domain. A second-order central difference scheme is employed for the vorticity transport equation. The vorticity field is updated at each time step via the alternative direction implicit technique \cite{birkhoff1962alternating}. The Poisson equation is solved with a semi-spectral method, wherein the Fourier transformation is employed along the azimuthal direction, and the second-order finite-difference method is used along the radial direction.

The proposed procedure can successfully capture the nonlinear interaction between the fluid and structure. The model has been validated with several canonical problems in Wang and Shoele 2021 \cite{wang2021geometrically}. In appendix \S\ref{apx:grid}, we present the grid refinement study for the current research.

\subsection{\label{subsec:mPOD}Geometrically Weighted Multi-Scale Proper Orthogonal Decomposition}
The FSI response of the foil and flap system involves different time scales, including the flap oscillation, the plunging foil, and the vortex shedding. To analyze this system and compare different mechanisms, we use the multi-scale POD \cite{mendez2019multi} with geometrical weighting \cite{wang2021geometrically}. This section will summarize the main components of the method.

Geometrically weighted modal analysis (GW-MD) obtains weighting functions through the Jacobian of the conformal mapping, which combines the structural motion and flow structure into a single spatio-temporal basis. GW-MD permits the use of data-driven modal analysis methods for FSI systems with deforming bodies. The procedure conserves the modal energy content and reveals how the structure and flow interact. To identify dominant flow modes for active flap cases, GW-MD is extended to allow multi-scale modal calculation. In particular, Multi-scale POD (mPOD) is applied in which a filter bank enables the splitting of the frequency spectrum into different scales. The frequency spectrum is preserved within each scale, and the corresponding optimal orthogonal eigenbases are assembled into a single mPOD basis. The following steps are done to perform geometrically weighted multi-scale POD (GW-mPOD):

\begin{enumerate}
	\item Selecting the flow data: in this study, we use vorticity to realize how the coherent structures form and dissipate. The $n$ snapshots of dimension $ n_r\times n_s$ are rearranged into column vectors of $\mathbf{u} \in \mathbb{R}^{n_r n_s \times 1}$ and assembled to form the data matrix $\mathbf{U} \in \mathbb{R}^{n_r n_s \times n}$.
	
	\item Computing the correlation matrix: the correlation matrix is calculated using $\mathbf{C} = (\mathbf{J} \odot \mathbf{U})^+ (\mathbf{J} \odot \mathbf{U})$ wherein the Jacobian is used as the weighting function ($\mathbf{J} \in \mathbb{R}^{n_r n_s \times n}$). Here, $\odot$ denotes Hadamard product defined as $(A\,\odot\,B)_{ij}\,=\,A_{ij}B_{ij}$ with $i$ and $j$ being free indices.
	
	\item Construction of the filter bank: the filter bank is formed to contain $m$ filters, $\mathbf{H_1}, ..., \mathbf{H_m}$. It is used to split $\mathbf{C}$ into $m$ non-overlapping contributions. \blue{The filter bank, in this research, is determined from the discrete Fourier transform of the correlation matrix. The filter bank consists of a low-pass filter, a high-pass filter, and a series of band-pass filters. The cutoff frequency of each band-pass filter is selected so that each filter contains an isolated spectral peak of the correlation matrix. The filter bank spams the entire spectrum, so the spectral content of the data set is retained through the partition. The data matrix is filtered in the spectral domain and transformed back to the temporal domain. POD is then performed on each temporal partition. An example process of applying a filter bank will be shown later in \S\ref{sec:competition} and readers are referred to the pioneering paper \cite{mendez2019multi} for mathematical details.
}
	
	\item Decomposing the correlation data matrix: Using the $m$ filters, the correlation matrix can be decomposed as $\mathbf{C} \approx \sum_{i=1}^{m}\bm{\Psi}_i \bm{\sigma}_i^2 \bm{\Psi}_i^T$, where $\bm{\Psi}$ is the temporal structure. Moreover, with the enforcement of the orthogonality condition, the spatial structures can be written as $\bm{\Phi} = \mathbf{U}\bm{\Psi}\bm{\sigma}^{-1} $. The spatial structures $\bm{\Phi}$ are sorted in descending order based on their contribution to the energy content.
\end{enumerate}

GW-mPOD has several unique benefits over alternative modal analysis methods for the current study. POD modes often suffer from spectral mixing. While dynamic mode decomposition (DMD) can overcome this and produce modes with a single frequency, those modes are not energetically relevant and not orthogonal. mPOD finds coherent structures with similar energy contents to POD modes, but at the same time, it reduces the spectral mixing problem. When combined with the geometrical weighting method, the analysis method can identify coherent flow structures related to the structure motion and internal flow dynamics. The identified flow features can then be compared based on their contribution to the energy content. The results will be demonstrated in \S\ref{sec:competition}.

\subsection{\label{subsec:setup}Problem Setup and Summary}
The parameter space of the FSI system in the current study includes the structural parameters, fluid properties, and active flap parameters. We focus primarily on how the active actuation of the flap affects the flow-induced plunging response of the foil. Sinusoidal oscillatory motions with different frequencies and amplitudes are imposed on the flap with a particular average angular position. Simulation parameters are defined as:
\begin{enumerate}
	\item \underline{Flow parameters}: The computational domain is extended to $15L$ around the foil and flap system with $L$ being the chord length of the foil. A grid with a resolution of $(280\times256)$ in radial and angular directions, respectively, is applied. The ambient flow velocity $U_\infty$ is chosen as the characteristic velocity scale, and the foil chord length $L$ is used as the length scale. Following previous literature \cite{guglielmini2004propulsive, menon2019flow}, the Reynolds number based on the chord length is fixed at $Re=\frac{U_\infty\,L}{\nu}=1000$. This Reynolds number is sufficiently large to allow the complex vortex shedding that drives the fluttering phenomenon. 
	\item \underline{Physical parameters of the foil}: \blue{The mass ratio of the foil is set to $m\equiv \frac{m_f}{\rho L^2}=0.1$.} The rotational degree of freedom of the foil is fixed and only the heaving motion is permitted. The spring-damper system is considered with the non-dimensional stiffness of $\frac{k_h}{\rho U_{\infty}^2}=1.0$ and Rayleigh stiffness proportional damping of $c_h=\lambda_c k_h$ with  $\lambda_c=0.1 \frac{U_{\infty}}{L}$. The flap length $l$ is selected based on the NASA design \cite{van2002aerodynamic} as $0.12L$. The effect of different inherent structural stiffness and damping have on the flow induced motion of representative cases is discussed in appendix \S\ref{apx:stiffness}. 
	\item \underline{Control parameters of foil}: A periodic pitching motion is imposed on the flap as $\theta=\theta_0+A\,\sin(\Omega\,t)$, where $\theta_0$ and $A$ are the mean angle and the amplitude of the flap relative to the chord respectively, and $f=\frac{\Omega}{2\pi}$ is the frequency of the flap oscillation. In the rest of this paper, we use the flap's Strouhal number, $St_f=\frac{f\,L}{U_\infty}$, to refer to the actuation frequency.
\end{enumerate}

\section{\label{sec:qualitative}Flow-induced Response of a Foil with an Active Flap}
The plunging response of a heaving foil subject to ambient flow and forced morphing flap motion is discussed in this section. Representative cases are inspected to identify the reasons behind different dynamic responses. Fourier transforms are employed to find the characteristic frequencies of the heaving motion. We identify how the motion evolves and how the trajectory of the foil motion changes in each period with the help of the phase portrait plot, a mapping between the heaving displacement $h$ and heaving velocity $dh/dt$. In addition to the structural response, the coherent structures associated with the dominant frequencies are examined with GW-mPOD. We will primarily consider a small angle of attack of AoA$\le10^\circ$ as this is within common operating conditions.

\begin{figure}
	\centering
	\makebox[\textwidth][c]{\includegraphics{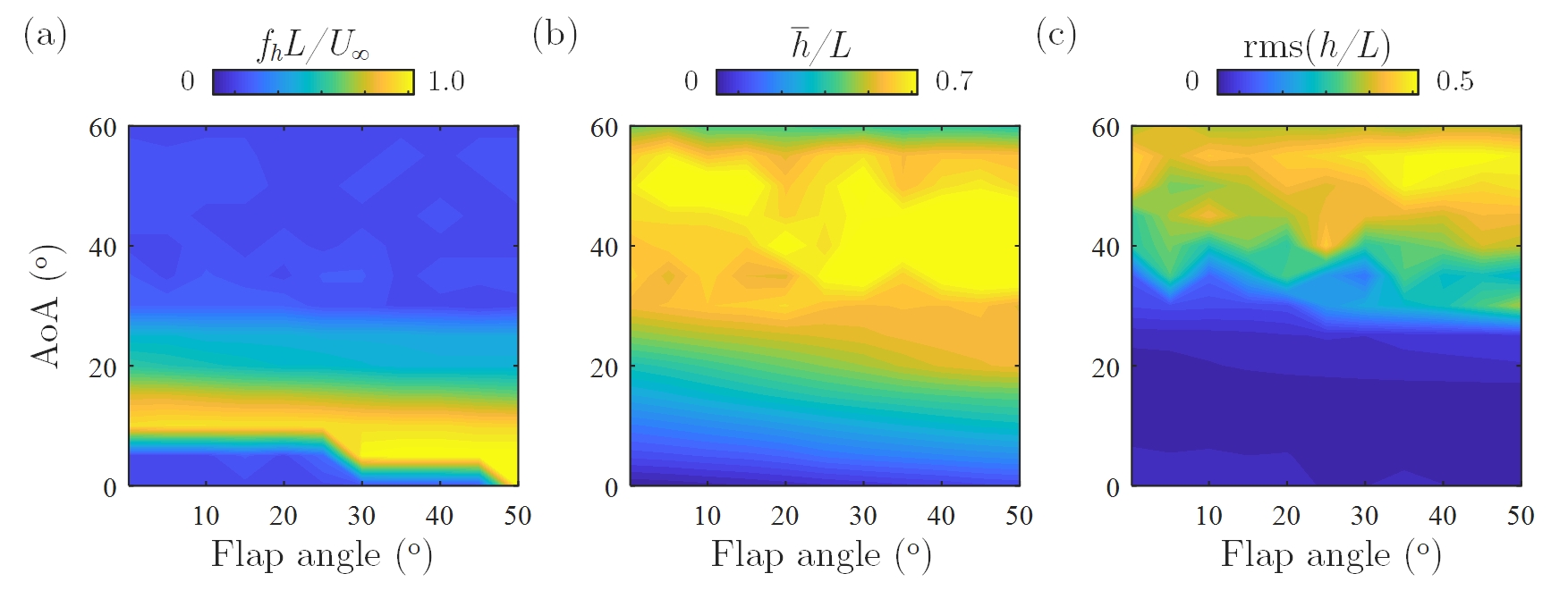}}
	\caption{(a) Dominant frequency of the heaving motion, (b) mean heaving displacement, and (c) rms of heaving amplitude of the uncontrolled cases.}
	\label{fig:2}
\end{figure}

\blue{
As a baseline, we consider the uncontrolled case where the flap is fixed at a specified AoA. Figure \ref{fig:2}(a) displays the dominant heaving frequency of the uncontrolled cases. The dominant frequency is highly dependent on the AoA. At low AoA, the flow is attached, and the airfoil has a very small mean displacement and oscillation amplitude, as shown in Fig. \ref{fig:2}(b-c). When the flow is attached, the added mass tensor coefficient for a NACA0012 airfoil is $a_{22} \sim 0.79$ \cite{la2012added} as we fixed the movement to only vertical plunging. The analytical natural frequency can be approximated as $f_0 = \frac{1}{2\pi} \sqrt{\frac{k}{m_f+m_a}}\sim0.19$, where $k=1.0$ is the nondimensional stiffness and the total virtual mass $m$ combining foil mass $m_f$ and added mass $m_a$ is $m=m_f+m_a=0.1+a_{22} \pi \rho  0.5^2 \sim 0.72$. The calculated natural frequency is very similar to what we observed in Fig. \ref{fig:2}(a) where the heaving frequency falls in the range of 0.16 when the flow remains attached when AoA$=0$. In the moderate AoA range, the continuous vortex shedding induces vortex-induced vibration, still with a small amplitude. In these cases, a narrow-band frequency response appears due to the alternative vortex shedding from the leading and trailing airfoil edges. The vortex shedding forms a regular vortex street in the wake and induces a steady periodic heaving motion. Interestingly, for the flap angles higher than a certain threshold, the apparent foil shape is modified and the periodic vortex shedding shifts to higher AoA. This response switch occurs at an average flap angle of $\theta \sim 30^\circ$. Large mean heaving displacement and oscillation amplitude are observed at higher AoAs due to the non-periodic vortex shedding from the leading and trailing edges and the airfoil stall at AoA$\sim 50^\circ$. The flap angle is not a decisive factor in this regime. Instead, the vortex-induced vibration is primarily affected by the AoA. 
}

\begin{figure}
	\centering
	\makebox[\textwidth][c]{\includegraphics{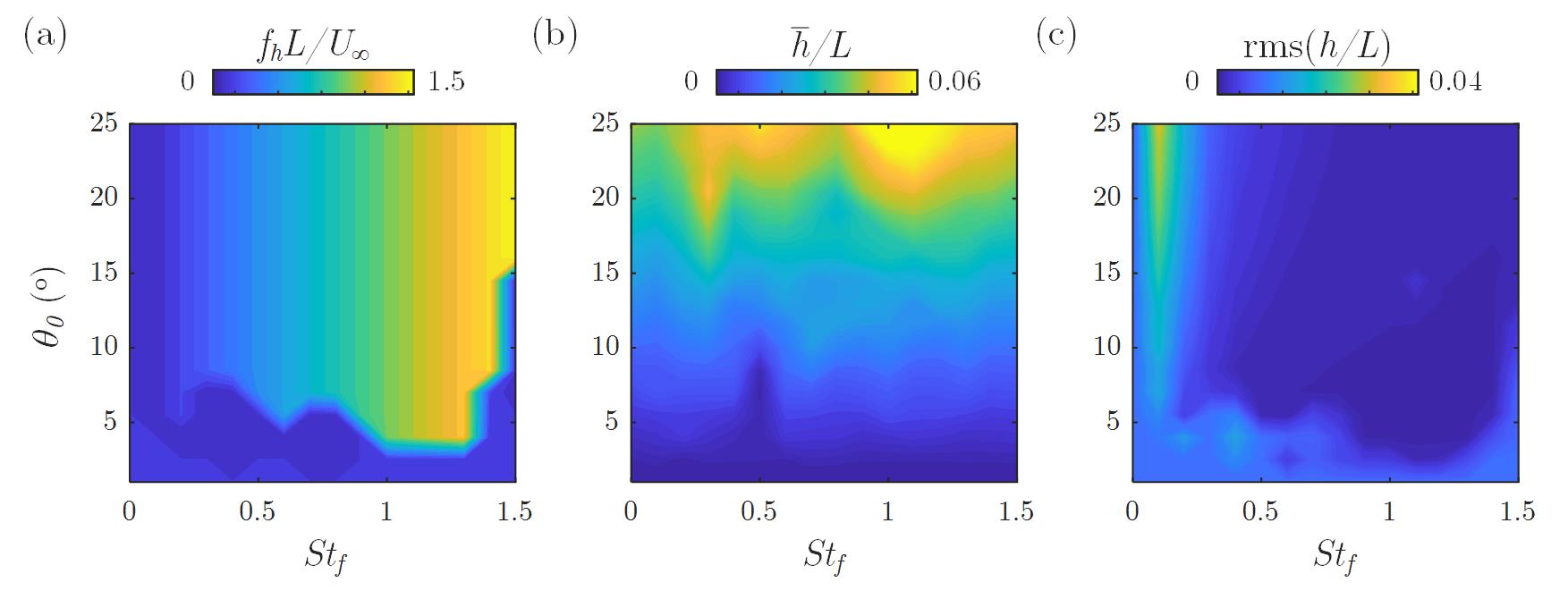}}
	\caption{(a) Dominant non-dimensional heaving frequency, (b) mean heaving displacement, and (c) unsteady heaving amplitude (rms) for AoA$=0^\circ$ and different flap positions $\theta_0$ and flap frequency $St_f$. Flap amplitude is fixed at $A/\theta_0=1$.  }
	\label{fig:3}
\end{figure}

Figure \ref{fig:3} shows how the foil heaving frequency $f_h$, average heaving displacement $\overline{h}/L$ and unsteady heaving motion of the foil varies with mean flap angle and flap frequency when the foil is at AoA$=0^\circ$. Here, it is assumed that $A/\theta_0=1$. When the amplitude is small, the flap motion does not lead to the plunging motion of the foil. At larger amplitudes, on the other hand, the foil starts to vibrate at the exact flap frequency. No significant change in vortex shedding pattern is observed even at larger flap amplitudes and the heaving displacement remains very small. 
\blue{
From the uncontrolled case, we can extract the average heaving displacement at AoA$=0^\circ$ and compare the difference between flap angle $\theta=0^\circ$ and $\theta=A$. It is seen that this value matches with the heaving amplitude in the active flap case with flap frequency $St_f=0.1$. This implies that when the flap is moving slowly, the change of the lift coefficient determines the heaving motion. However, when the flap frequency increases, we see a very small heaving amplitude indicating almost no vibration. What happens is that the flap moves too fast and the induced motion from the change of lift is damped out. The cause of the small heaving motion at higher flap frequency, with the same frequency as the flap motion, is the added mass effect due to the flap motion.}

\begin{figure}
	\centering
	\makebox[\textwidth][c]{\includegraphics{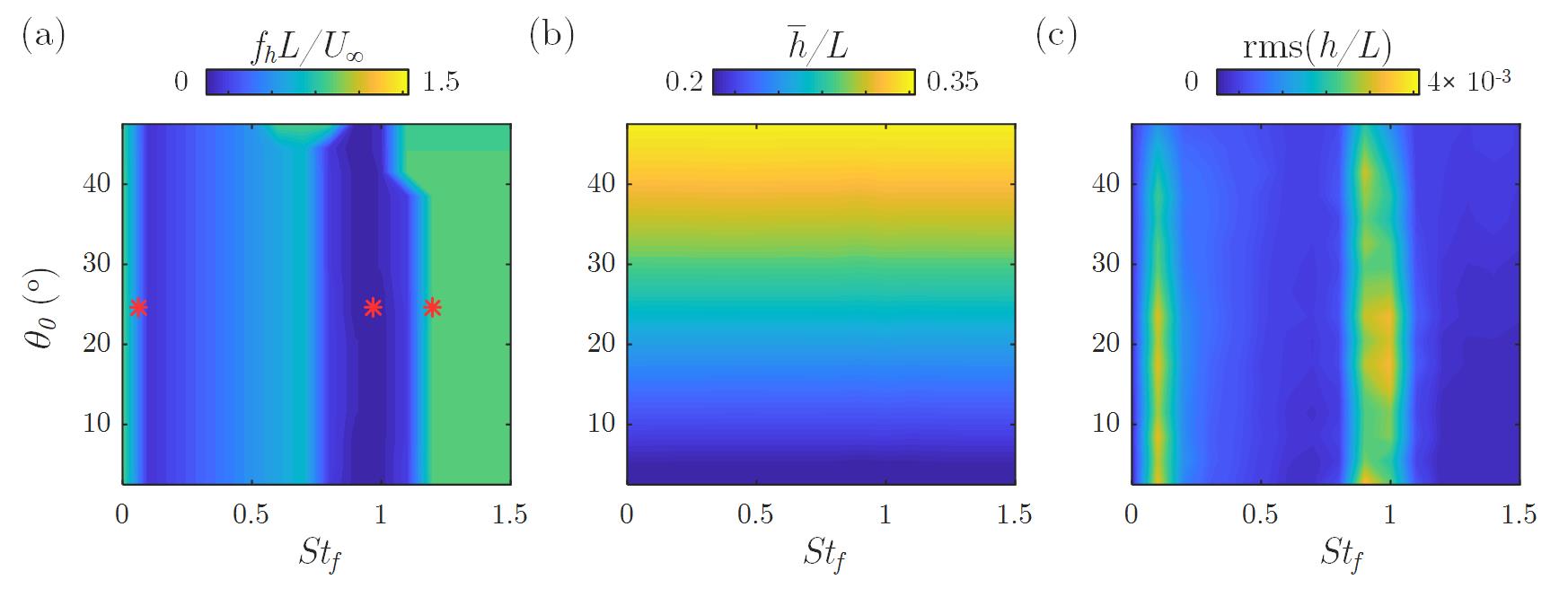}}
	\caption{(a) Dominant non-dimensional heaving frequency, (b) average heaving displacement, and (c) heaving amplitude (rms) for AoA$=10^\circ$ with flap amplitude $A=1.5^\circ$ at different average angle $\theta_0$.}
	\label{fig:4}
\end{figure}

An increase in AoA prompts substantial changes. Figure \ref{fig:4} plots the heaving response with AoA$=10^\circ$ and with a very small flap amplitude of $A=1.5^\circ$. At low flap frequencies, the foil heaves at the flap frequency, while at higher flap frequencies, the foil switches its dynamic response and instead oscillates at a terminal fixed frequency. Interestingly, a low-frequency band around the natural vortex-induced heaving frequency ($f_h L/U_{\infty}\sim0.96$) separates these two dynamic regions. From the amplitude plot, we observe that this low-frequency band coincides with the large vibration amplitude. We will take a deeper look at this region later. Besides, similar to the zero AoA case, a narrow large heaving amplitude band is located at $St_f\sim0.1$. \blue{This amplitude is similar to the difference of the heaving displacement with different static flap angle (uncontrolled cases). The slow flap movement changes the lift coefficient, which causes a large amplitude vibration. At higher flap frequencies, this effect is damped out, but the flap-induced vortex street creates another large-amplitude peak as will be explained.} The mean heaving displacement is only proportional to the mean flap angle, whereas the larger mean flap deflection results in a higher lift force on the foil and larger mean displacement. 

\begin{figure*}[!]
	\centering
	\makebox[\textwidth][c]{\includegraphics[width=1.0\textwidth]{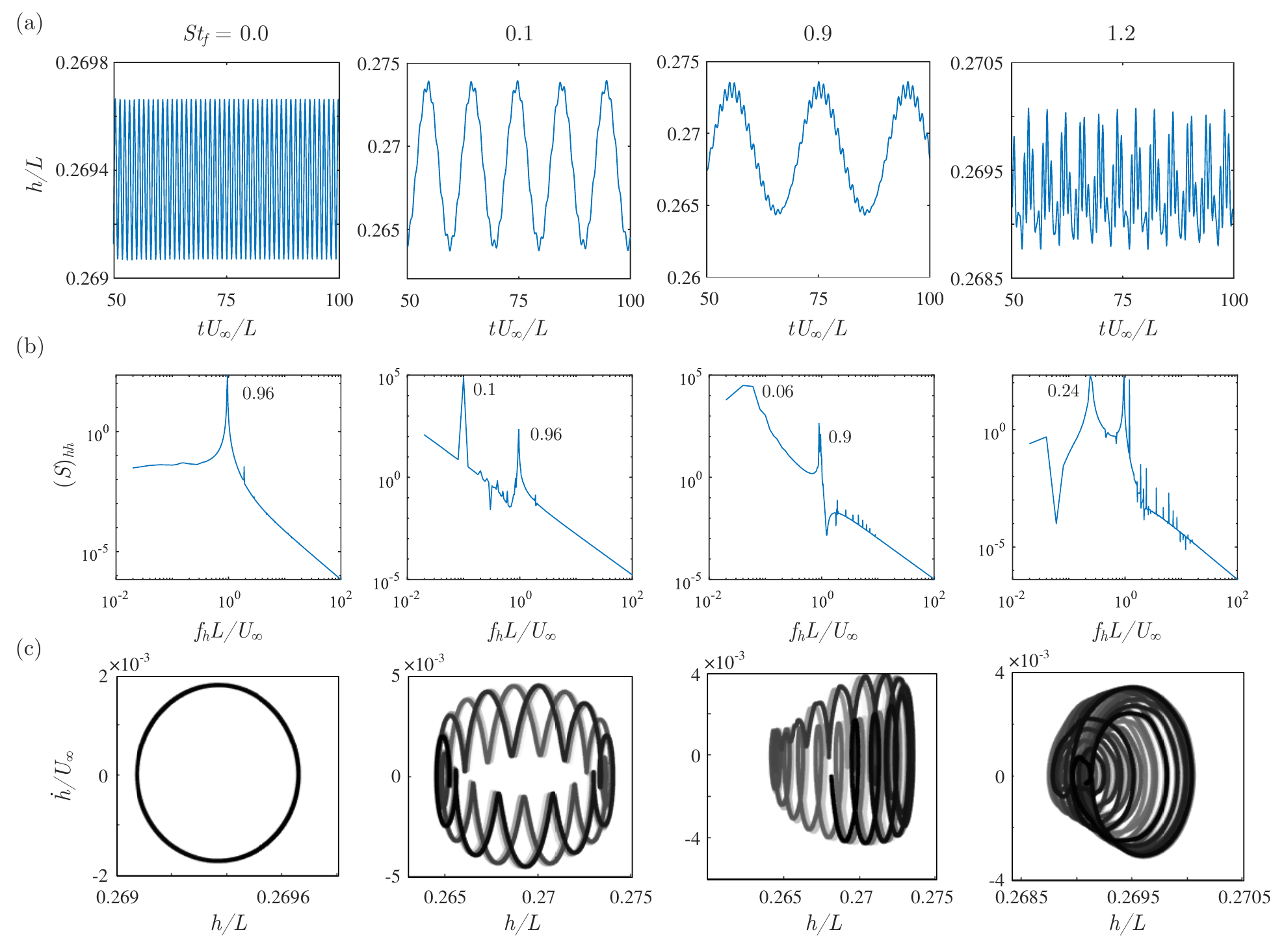}}
	\caption{The (a) heaving motion time histories, (b) power spectra, and (c) the phase portraits of $St_f={0.0,\, 0.1,\, 0.9,\, 1.2}$. Here, it is assumed that AoA$=10^\circ$, $\theta_0=24^\circ$ and $A=1.5^\circ$.}
	\label{fig:5}
\end{figure*}

To further explain the characteristic dynamics of these two regions, three representative cases are discussed below. All of these three active flap cases have the same AoA of $10^\circ$, flap average position $\theta_0=24^\circ$ and flap amplitude $A=1.5^\circ$, but with different flap frequencies of $St_f=\{0.1,\,0.9,\,1.2\}$. The selected cases are also marked in Fig. \ref{fig:4} (a). Figure \ref{fig:5} shows the heaving displacement over a certain time interval, the power spectrum of the heaving displacement calculated from 10 flapping cycles, and the phase portrait of these cases. The phase portrait trajectories are marked with white to black color based on its temporal state. The uncontrolled case $St_f=0.0$ is also included for cross-comparison. 

\begin{figure}
	\centering
	\includegraphics
	{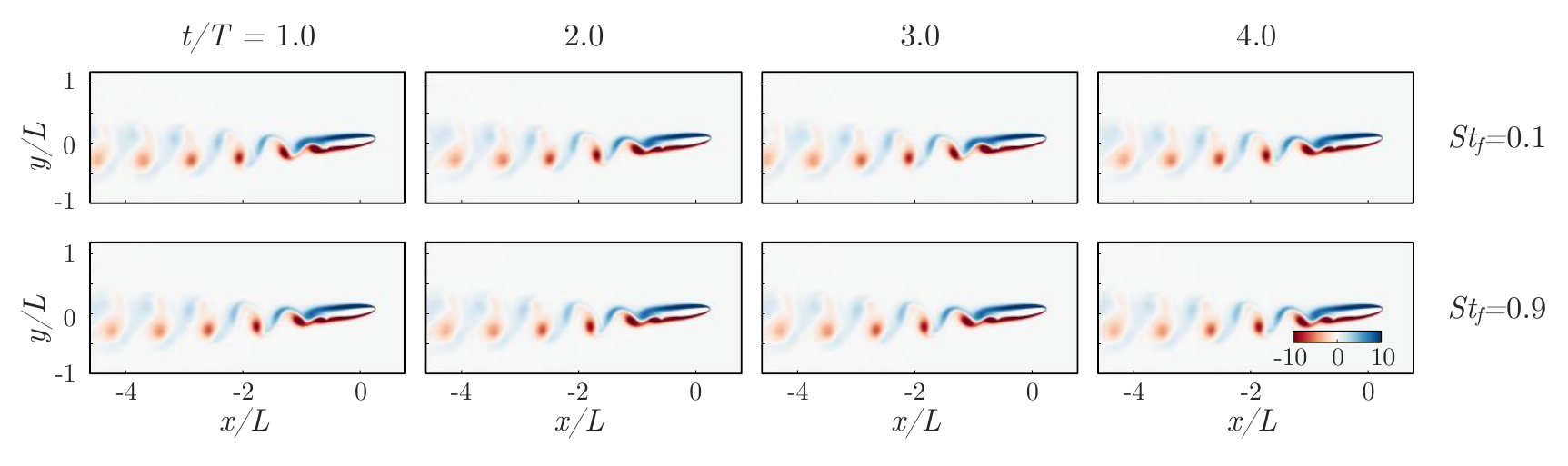}
	\caption{Instantaneous vorticity fields over four active flap periods of $St_f=0.1$ and $St_f=0.9$. $T$ is the flap motion period.}
	\label{fig:6}
\end{figure}

The uncontrolled case with no flap motion undergoes a periodic oscillation with a frequency of $0.96$ due to the vortex-induced vibration. This response mode is called the flow-induced mode from now on. When the actuation is activated, we can see that torus-like trajectories emerge for all three selected cases. The heaving time history involves large-scale oscillation and small amplitude unsteady fluctuation. Furthermore, the spectrum clearly shows multiple distinct peaks indicating that the system is quasi-periodic and the observed dynamics are from mode competition between different mechanisms. In particular, $St_f=0.1$ case has distinct peaks at the flap frequency $0.1$ and flow-induced frequency $0.96$, while the other case with $St_f=0.9$ shows two peaks at the frequencies of $0.9$ and $0.06$. Although they share a very similar heaving response and flow field (shown in Fig. \ref{fig:6}), in the previous study \cite{wang2021geometrically} we highlighted that different mechanisms are involved in these two cases.  \blue{The results will later be ascertained here using the GW-mPOD analysis with extra knowledge on the energy content and the competition between the modes.} In the low flap frequencies, the flap motion does not substantially modify the downstream wake and mainly adjusts the exit angle of the vortex wake. On the other hand, the flap motion of $St_f=0.9$ creates a wake structure that interferes with the flow-induced mode. The other case of $St_f=1.2$ exhibits much smaller heaving amplitude similar to the non-actuated case and have a more heaving movement closer to quasi-periodic motion. 

\begin{figure}
	\centering
	\makebox[\textwidth][c]{\includegraphics{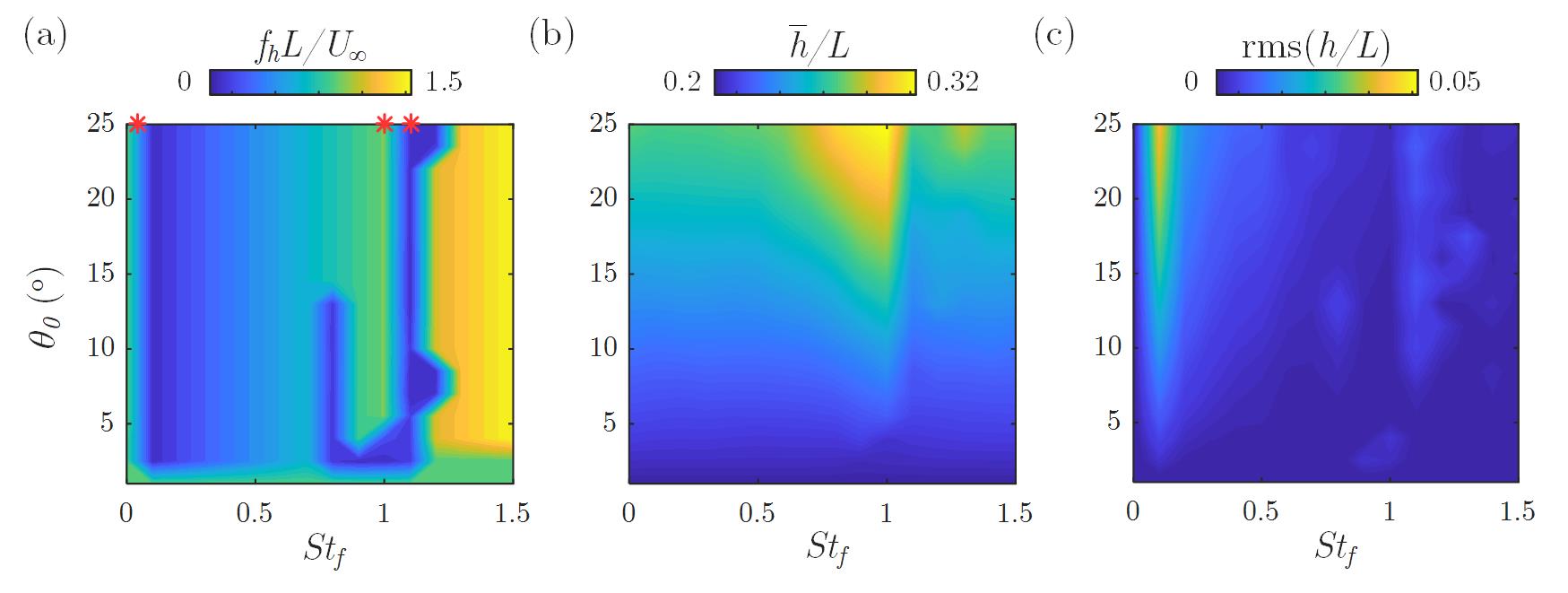}}
	\caption{(a) Dominant non-dimensional heaving frequency, (b) mean heaving displacement, and (c) unsteady heaving amplitude (rms) for AoA$=10^\circ$ with the large amplitude flap motion of $A/\theta_0\,=\,0$.}
	\label{fig:7}
\end{figure}

A very different foil-fluid interaction response is observed for the flap with a large oscillation amplitude of $A=\theta_0$ (Fig. \ref{fig:7}). Here, the low-frequency band narrows while the foil still undergoes a large heaving motion at the flap frequency outside of this band. Interestingly, there are two branches of the low-frequency band. Also, the average heaving displacement does not increase monotonically with the average flap angle in the case of the small flap amplitude. Instead, it shows a unique peak near the natural frequency range. The heaving amplitude peak at the slow flap frequency emerges again for the same reason as before: the slow transition of the overall lift coefficient.

\begin{figure*}
	\centering
	\makebox[\textwidth][c]{\includegraphics[width=1.0\textwidth]{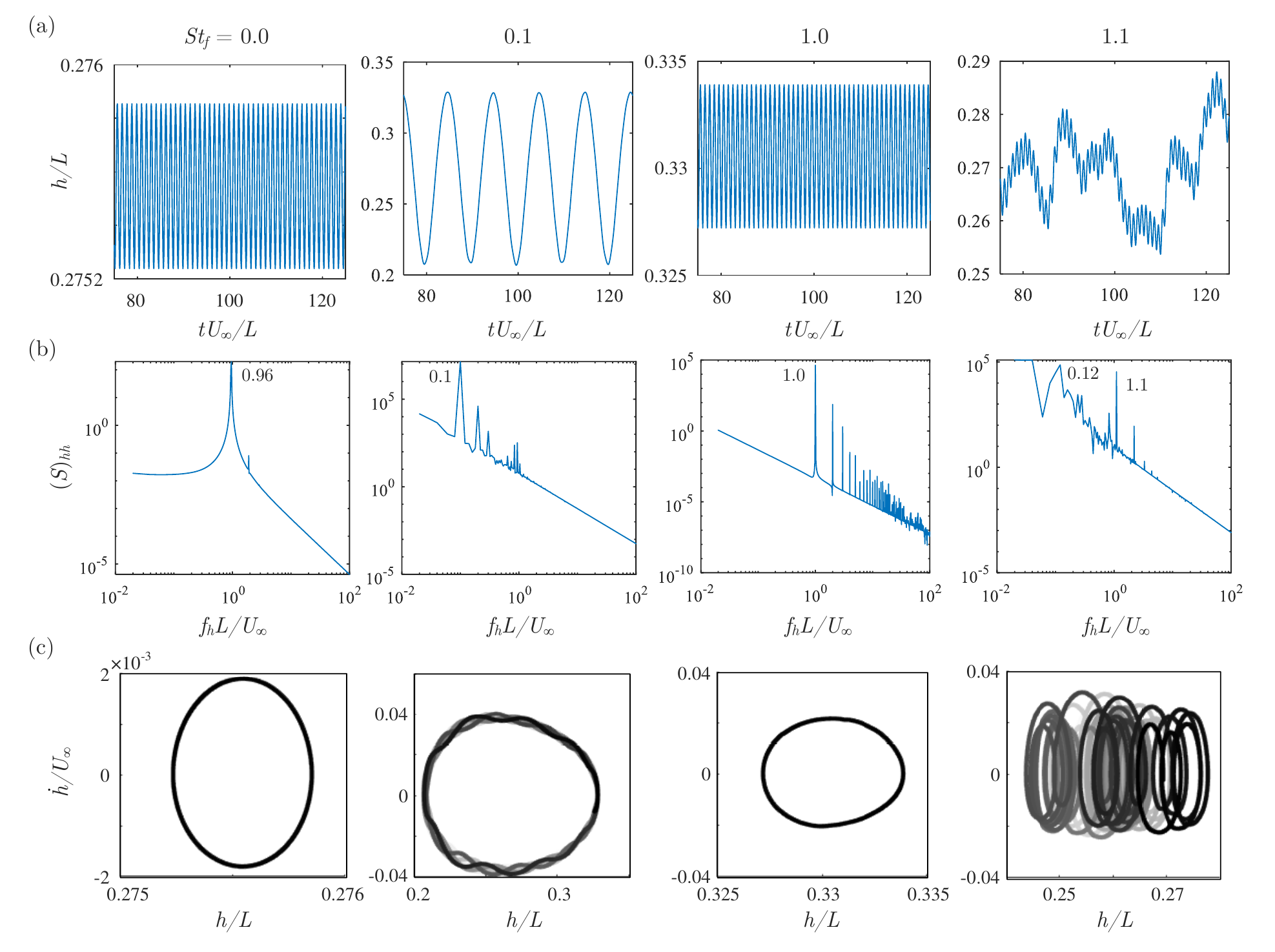}}
	\caption{The heaving displacement time history, power spectrum, and the phase portrait plot of $St_f={0.0,\, 0.1,\, 1.0,\, 1.1}$, respectively from the top to bottom row. The results are for AoA$=10^\circ$ and $A=\theta_0=24^\circ$.}
	\label{fig:8}
\end{figure*}

The time history, power spectrum, and the phase portrait of the heaving motion are shown in Fig. \ref{fig:8}. Rows are associated with different flap actuation frequencies of $St_f\,=\,\{0,\, 0.1,\, 1.0,\, 1.1\}$. When $St_f=0.1$, the larger flap oscillation amplitude creates a stronger pressure gradient over the flap. During the downsweep, the pressure gradient creates a new vortex behind the trailing edge. The newly formed vortical structure then separates from the flap during the upsweep and merges into the wake. The leading-edge vortex shedding is locked onto the same period due to the large suction pressure caused by the flap, which results in the purely periodic motion at the flap frequency $St_f=1.0$. This can be identified from the clear resonance peaks in the power spectrum and periodic orbit in the phase portrait. A similar lock-in phenomenon has been reported in several recent studies of airfoils with morphing surfaces \cite{lei2014unsteady, jones2018control, kang2020lock}. 

\begin{figure}
	\centering
	\makebox[\textwidth][c]{\includegraphics[width=1.0\textwidth]{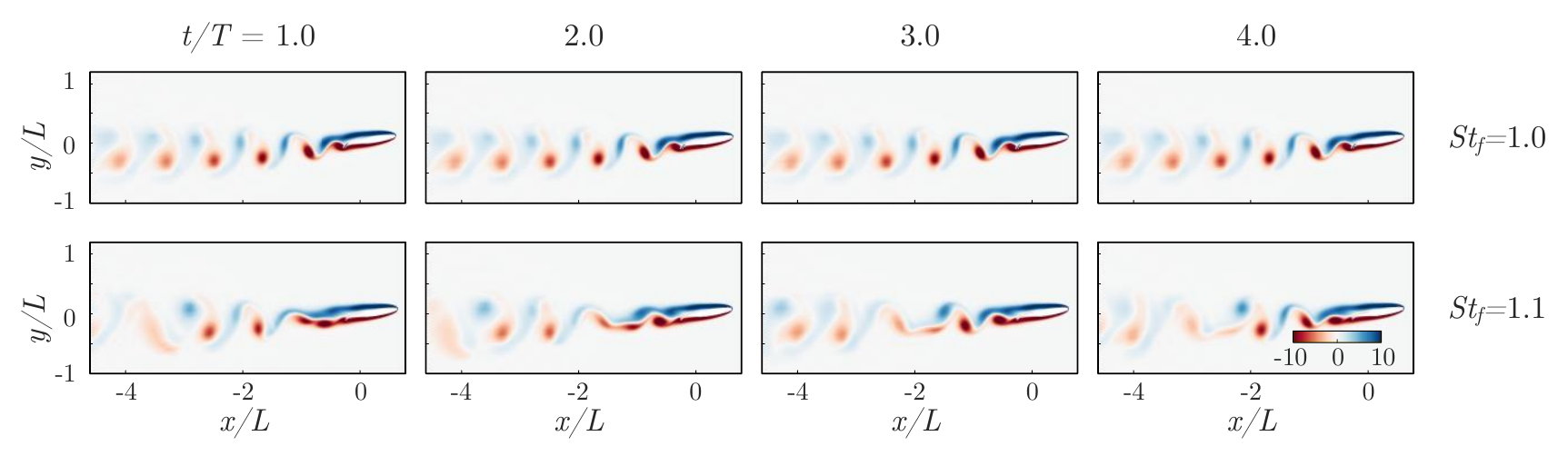}}
	\caption{Instantaneous vorticity field over four active flap periods of (a) $St_f=1.0$ and (b) $St_f=1.1$ with AoA$=10^\circ$ and flap motion $A=\theta_0=24^\circ$.}
	\label{fig:9}
\end{figure}

When the flap frequency increases to $St_f=1.1$, the vortex shedding from the trailing edge becomes chaotic (Fig. \ref{fig:9}). The leading-edge vortex shedding mode is now controlled by the foil geometry and its natural frequency, while at the trailing edge, the flow separation is dictated by the flap motion. Since there is a discrepancy between the flap frequency and natural vortex shedding at the trailing edge, an intermittent wake structure is formed. The heaving motion is consequently chaotic, as demonstrated by the never-repeating trajectory in the phase portrait.

The large heaving amplitude of $St_f=0.1$ originates from the low-frequency large lift forces modifications caused by the continual geometrical changes. When the flap slowly changes its angle, it produces different lift forces at different vibration stages. Consequently, the fluid-structure interaction of the foil is quasi-static, where the flap movement does not create a new vortical structure, but it still affects the foil heaving motion by modifying the instantaneous lift force. This is supported by the presence of a spectrum peak at flap frequency $f\,L/U_\infty=0.1$. This peak is present in all three cases discussed so far, and the mechanism behind it can be related to the uncontrolled cases.

\blue{In summary, the flap oscillating with a larger amplitude impacts the flow and heaving motion more than the smaller amplitude case. The larger flap amplitude substantially modifies the flow at the flap location and plays a controlling role in determining system response than the smaller amplitude cases.} When the flap frequency approaches the natural frequency, there is a lock-in condition and the foil heaves at the same frequency as the flap movement. On the other hand, when the flap frequency increases further, there is substantial vortex shedding from the flap and foil surfaces resulting in a chaotic heaving motion of the foil. Finally, at a specific low frequency, the slowly changing geometry leads to quasi-static relocation of the foil through a readjustment of the flow. After observing such a diverse interaction between the flap-induced and flow-induced modes, a question arises: is there an index that can classify the competition between these two characteristic modes and represent the system behavior? This is achieved using the GW-mPOD analysis in the next section.

\section{\label{sec:competition}Mode Competition between Flap-induced and Flow-induced Modes}
We now apply the GW-mPOD to the observed vorticity field to investigate the interaction between different fluid dynamic modes that drive the foil's heaving motion. More than 30 flap oscillation cycles with at least 10 snapshots per cycle are used to perform the modal analysis process for all the following cases.  

\begin{figure}
	\centering
	\makebox[\textwidth][c]{\includegraphics[width=1.0\textwidth]{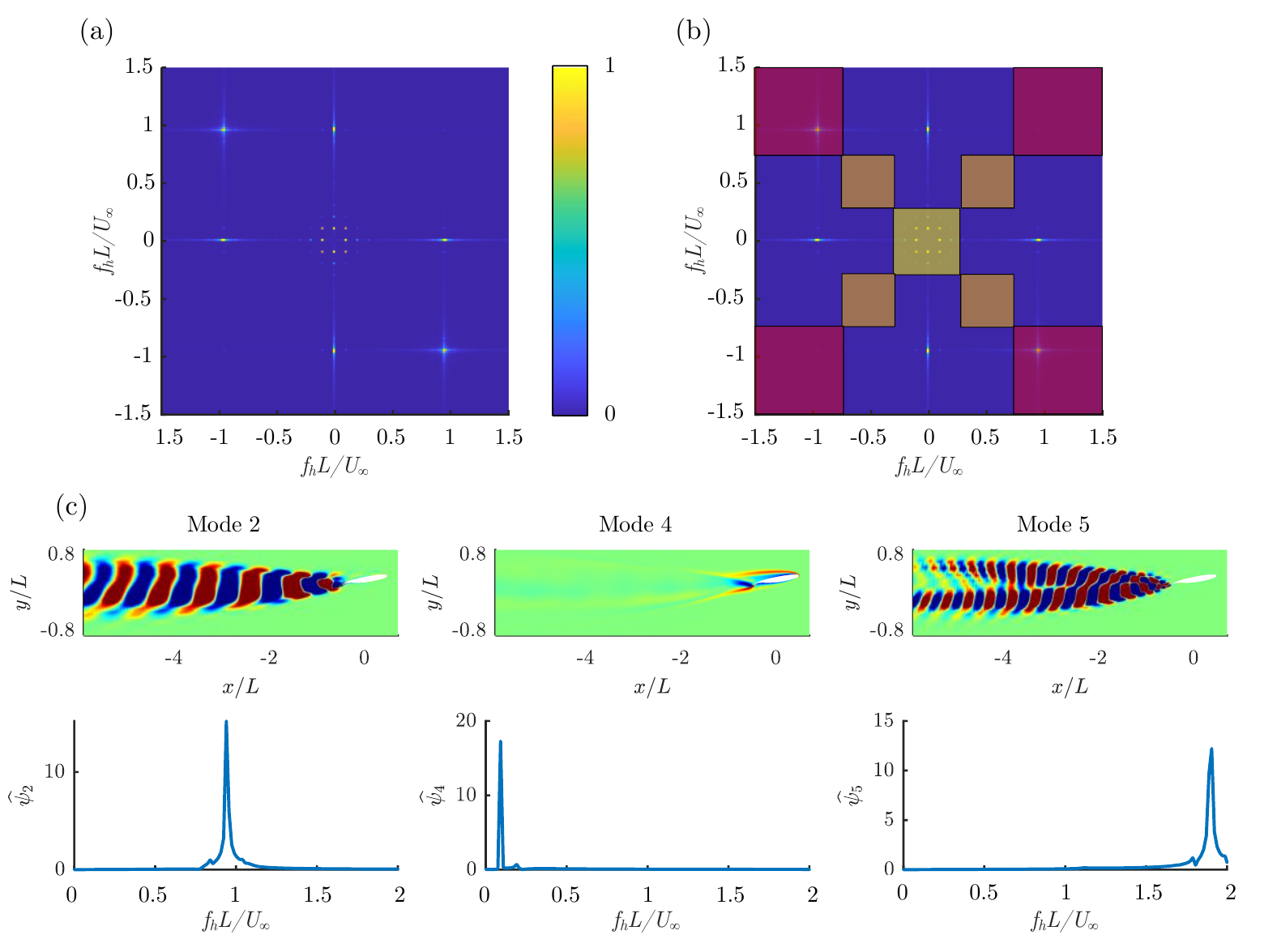}}
	\caption{(a) The discrete Fourier transform (DFT) of the normalized temporal correlation matrix, (b) partitioned correlation matrix. (c) shows the spatial structure and spectrum of the leading modes of the case AoA$=10^\circ$ with flap frequency $St_f=0.1$, amplitude $A=1.5^\circ$ and average angle at $\theta_0=24^\circ$. \blue{The modes are normalized based on their respective largest vorticity, which is the same for follwing figures.}}
	\label{fig:10}
\end{figure}

Figure \ref{fig:10} demonstrates the GW-mPOD procedure for the foil-flap case of AoA$=10^\circ$ and $St_f=0.1$. Here, flap actuation is fixed at $A=1.5^\circ$ and $\theta_0=24^\circ$. \blue{Fig.\ref{fig:10}(a) shows the discrete Fourier transform of the geometrically weighted temporal correlation matrix, and based on the intensity (spectral power), the correlation matrix is partitioned with the filter bank as shown in Fig. \ref{fig:10}(b).} Conventional snapshot POD procedure \cite{sirovich1987turbulence} is then applied to each partition. The flow vorticity modes are sorted based on their enstrophy contribution. The leading modes all contain pure spectral contents and are associated with different dynamic responses.

\begin{figure}
	\centering
	\makebox[\textwidth][c]{\includegraphics[width=1.0\textwidth]{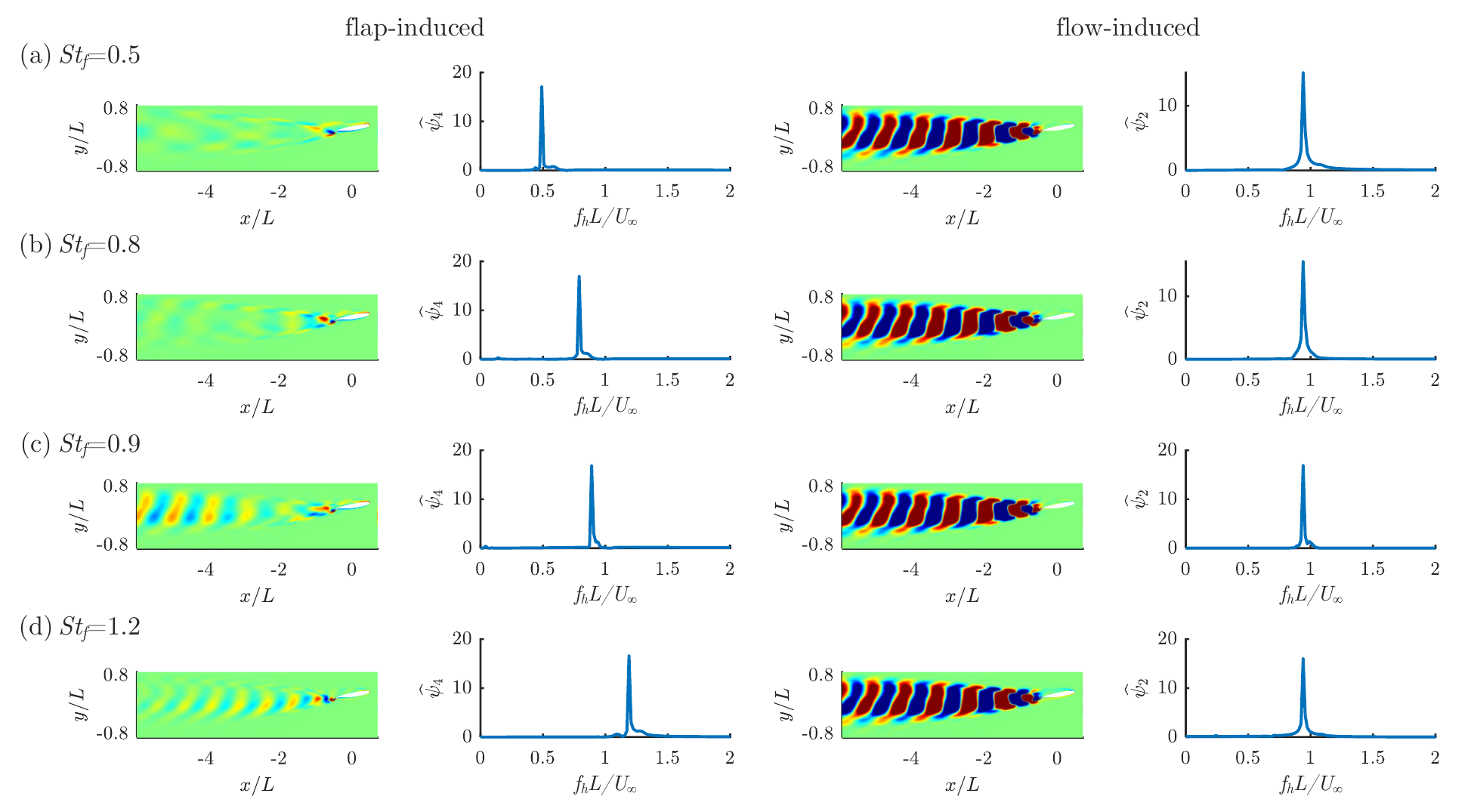}}
	\caption{The spatial structures and spectra of the leading flap-induced and flow-induced modes of the case AoA$=10^\circ$ with flap amplitude $A=1.5^\circ$, average angle $\theta_0=24^\circ$ and different flap frequency $St_f$.}
	\label{fig:11}
\end{figure}

The GW-mPOD is employed for representative cases identified in the previous section. The first case is with a small amplitude flap motion and $St_f=0.1$ (Fig. \ref{fig:10}(c)). Here, mode 2 has the same pattern and frequency as in the uncontrolled case, thereby it is called the flow-induced mode. On the other hand, mode 4 has the same frequency as the flap oscillation and is denoted as the flap-induced mode. We can see that there is little extended wake in this mode so it only affects the area close to the trailing edge. The trailing edge vorticity field shows two non-convective vortical structures that are formed through the slow flap sweeping motion. The slower motion does not create enough pressure change to force vortex shedding but can adjust the exit angle of the flow-induced wake at the trailing edge. This is why the foil undergoes quasi-periodic motion, with two distinctive modes superimpose upon each other. The slow modulation of the flap changes the direction of the steady vortex shedding, modifies the lift force and results in a low-frequency, high-amplitude heaving motion. Similar behavior can be observed for all flap frequencies $St_f\le0.8$ (Fig. \ref{fig:11}(a,b)). However, when the flap frequency reaches $St_f=0.9$ (Fig. \ref{fig:11}(c)), the flap-induced vortices propagate into the wake. At higher frequencies, the flap-induced mode has a clear extended wake as demonstrated for $St_f=1.2$ (Fig. \ref{fig:11}(d)). When the flapping frequency increases, there will be subsequent interference between the flap-induced and flow-induced modes. Here, a stronger pressure gradient from the flap motion induces a new convective wake structure and, consequently, the interference between two energetically similar wakes promotes quasi-periodic motion of the foil. 

\begin{figure}
	\centering
	\makebox[\textwidth][c]{\includegraphics[width=1.0\textwidth]{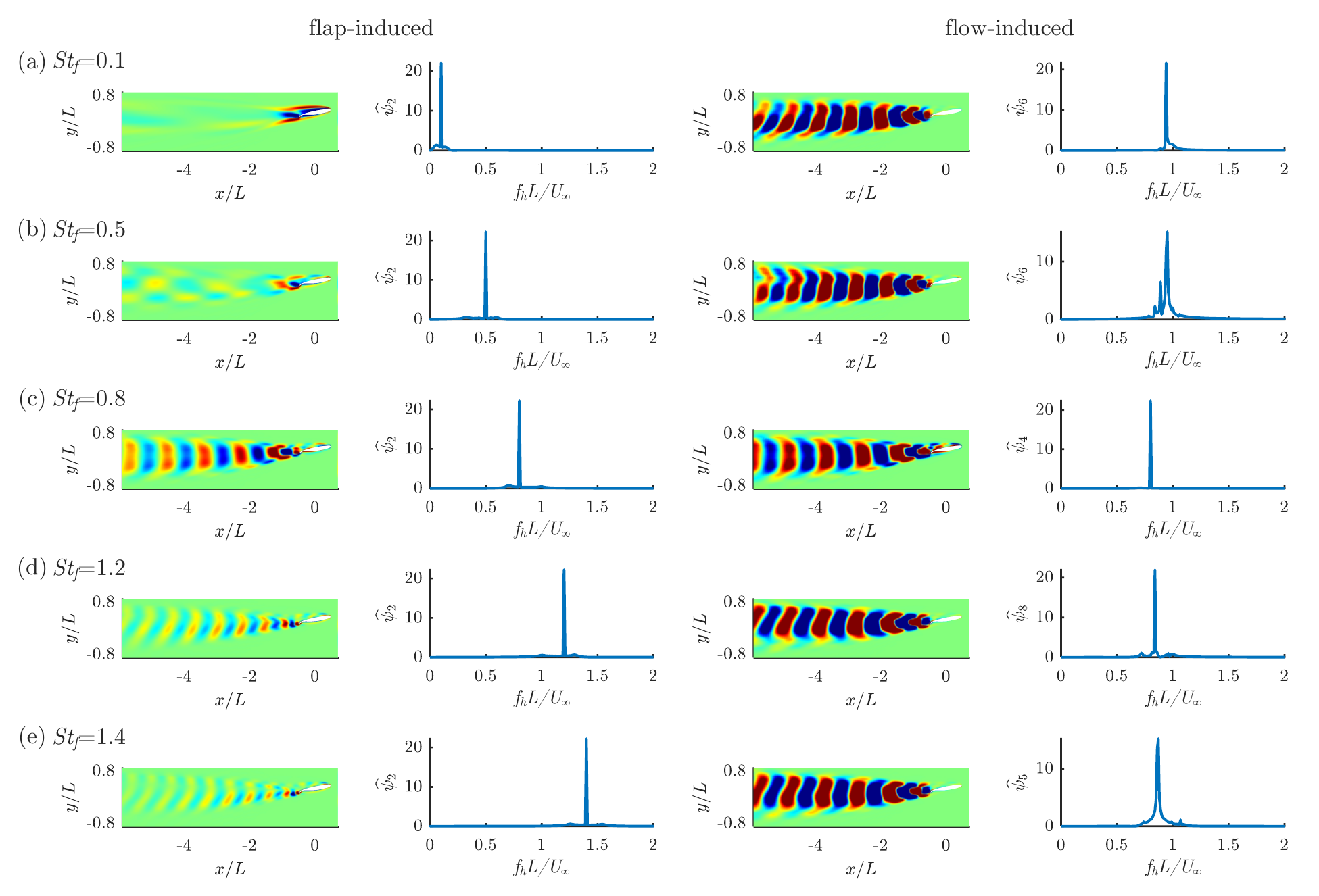}}
	\caption{The spatial structures and spectra of the flap-induced and flow-induced modes of the case AoA$=10^\circ$ with flap amplitude $A=\theta_0=24^\circ$ and different flap frequency $St_f$. The modes are normalized based on their respective largest vorticity.}
	\label{fig:12}
\end{figure}

The large flap oscillation amplitude of $A=\theta_0=24^\circ$ is considered next. Fig. \ref{fig:12} shows the leading flap-induced and flow-induced modes along with the frequency spectrum of the representative cases. We observed that $St_f=0.1$ (Fig. \ref{fig:12}(a)) exhibits large-amplitude, low-frequency plunging motion similar to the case with small flap amplitude. The flap-induced mode also shows a comparable spatial pattern, indicating that both cases are alike at modifying the exit angle of the trailing edge wake. However, the vorticity intensity is much stronger here, and the flap-induced mode contributes more to the enstrophy than the flow-induced mode. Once the flap frequency increases beyond $St_f=0.3$, an extended wake develops. At the higher flap frequency of $St_f=0.8$ (Fig. \ref{fig:12}(c)), the flap-induced and flow-induced modes share the exact same frequency and the system is in a perfect lock-in state. This is different from the small-amplitude case where the flow pattern is similar, but the frequencies of these modes are different. By increasing the flap frequency to $St_f=1.2$ (Fig. \ref{fig:12}(d)), the flap and flow-induced modes separate again.

\begin{figure}
	\centering
	\includegraphics 
	{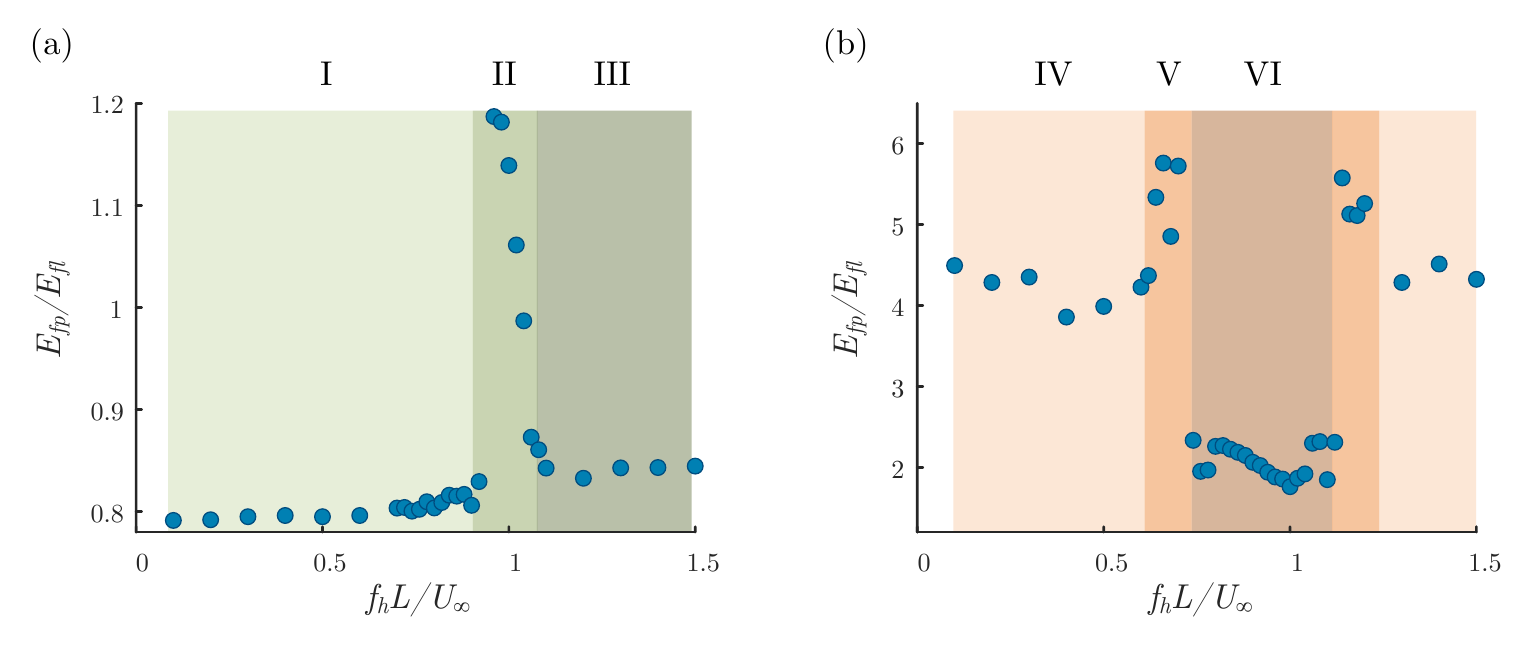}
	\caption{Ratio of the leading flap-induced and flow-induced modal energy of AoA$=10^\circ$ case with (a) small flap amplitude $A=1.5^\circ$ and average angle $\theta_0=24^\circ$ and (b) large amplitude $A=\theta_0=24^\circ$.}
	\label{fig:13}
\end{figure}

We calculated the modal enstrophy ratio between the leading flap-induced ($E_{fp}$) and flow-induced ($E_{fl}$) modes for the small and large flapping amplitude and different flapping frequencies (Fig. \ref{fig:13}). In both cases, the enstrophy ratio can be divided into three regions. First, we consider the small amplitude case in Fig. \ref{fig:13}(a). At low flap frequencies, $E_{fp}/E_{fl}\sim 0.8$, but it suddenly jumps to much larger values of $E_{fp}/E_{fl}\sim 1.2$ when flap oscillation approaches the flow-induced vibration frequency. The enstrophy ratio eventually drops gradually after $St_f>1$ to $E_{fp}/E_{fl}\sim 0.85$. At a much higher $St_f$ range, the ratio increases with a much milder slope. Although the plunging motion could be classified as quasi-periodic with their multiple distinct peaks in all cases, the three regions indicate three different physical mechanisms, denoted with Roman numbers in Fig. \ref{fig:13}. When the flap oscillation amplitude and frequency are both small (I), the flap-induced mode is limited to adjusting the exit angle of the flow-induced wake. At large flap frequency (III), the flap motion induces extended wake, which has higher enstrophy and interferes with the flow-induced wake. Finally, when the flap frequency approaches the flow-induced frequency (II), the two modes have comparable enstrophy and similar spatial structure, which indicates the lock-in phenomenon where the flow-induced mode transports enstrophy to the flap-induced mode and hence are correlated. Recall that in Fig. \ref{fig:4}(a) we also saw three different regions in the heaving frequency, and the reason behind that is now clear: the flow-induced and flap-induced modes, each have a different controlling role at different flap frequencies. 

The enstrophy ratio of the large amplitude case shown in Fig. \ref{fig:13}{b} reveals a different story compared to the small amplitude case. Now, the flap-induced mode is always more energetic than the flow-induced mode. Again, three distinct regions can be identified. The first region is associated with a large dissimilarity of flap and flow-induced modal frequencies and $E_{fp}/E_{fl}\sim 4$ (IV). In this region, the flap-induced mode and flow-induced modes co-exist, which leads to quasi-periodic motion. When the frequencies become closer, the enstrophy ratio increases to $E_{fp}/E_{fl} > 5$ (V), implying that the flap-induced mode overpowers the flow-induced mode. The flap-induced vortex shedding dominates the flow, but the slightly mismatched leading-edge vortex shedding results in an unsteady lift force that eventually destabilizes the heaving motion. This explains why there are two branches of low-frequency bands in Fig. \ref{fig:7}(a): the flap-induced mode is dominant and the flow-induced mode is not lock-in due to the larger frequency gap between them, leading to a chaotic foil heaving behavior. Within the locked in region we see that the enstrophy ratio drops to $E_{fp}/E_{fl} \sim 2$ (VI). Once locked in, the leading edge vortex shedding is now guided by the trailing edge flap motion and the two modes excite the same flow modes. In both small and large amplitude cases, we observe that the energy is transported from the more energetic to less energetic mode. Hence, the enstrophy ratio is closer to unity. The lock-in region is wider for larger flap amplitude cases due to the more energetic flap-induced mode.

We now summarize the findings from the GW-mPOD analysis. There are two competing modes: flap-induced and flow-induced modes. When the flap oscillation amplitude is small, the two modes have a similar energy level and co-exist to create quasi-periodic heaving motion. When the flap frequency is low, the flap quasi-statically modifies the exit angles of the flow-induced wake, while at higher flap  frequency, the flap induces an extended wake that interferes with the flow-induced mode. On the other hand, when the flap oscillation amplitude is large, the flap-induced mode contains much higher energy and is the dominant mechanism. Particularly, when the frequencies of the two modes approach each other, a strong lock-in effect is observed. The trailing edge motion guides the leading edge vortex shedding and the two competing modes share a similar energy content. \blue{The observations can also explain why in the smaller flap amplitude cases the flow fields look similar throughout different actuation (Fig. \ref{fig:6}) but in the larger flap amplitude cases flow changes significantly (Fig. \ref{fig:9}): in the larger flap amplitude case, the flap-induced modes has higher enstrophy and is able to strongly interfere with the flow-induced modes.} GW-mPOD successfully reveals how these two passive and active modes bring rich dynamics to the system. The modal enstrophy ratio between these modes is identified as a predictive indicator of the nature of plunging motion. 

\section{\label{sec:conclusion} Conclusion}
In this paper, we employed a strongly coupled FSI algorithm to study the flow around a deforming geometry. Different analysis methods, including a geometrically weighted multi-scale modal analysis technique, are adopted to dissect the flow field. We explored dominant dynamic interaction modes that could be useful for designing control strategies to regulate the flow-induced fluttering response. Deployment of the GW-mPOD method leads to valuable observations of this complex FSI system, summarized as follows: (1) when the flap amplitude and AoA of the foil are small, the flap and flow-induced modes are equally important and the interaction between them primarily creates quasi-periodic motion; (2) at larger flap amplitudes, the flap-induced vortex shedding dominates the foil heaving motion, causing the heaving motion to lock onto the flap oscillation frequency; (3) the ratio of the flap-induced and flow-induced GW-mPOD modal energy is a discriminative indicator of the flow field and flow-induced vibration of the system. For the AoA considered, the active control surface is an effective mechanism in modulating the flow and regulating the foil motion.

The proposed methodology utilizes multi-scale analysis that considers both flow and structural motion. It can be used to dissect the flow over other actively deforming bodies to differentiate the potential role of competing modes. The energy content of the competing modes can be used to evaluate the effectiveness of different active control methods and determine the optimal geometry morphing.

Although it is shown that the observed dynamic response is robust with respect to the structural stiffness and damping values, other aspects of the active flap-foil interaction such as flap size or foil geometry require further investigation before their roles on the geometry-induced fluttering phenomena could be thoroughly explained. Another improvement would be to expand the current study to 3D setup using differential geometry techniques. Quasi-conformal mapping techniques are proper candidates for shape recognition and surface mapping. Methods like the discrete conformal mapping (DCM) preserves the angle of the mesh while ensuring Laplacian conservation \cite{gotsman2003fundamentals, floater2005surface, li2007conformal}, which is required ingredients to extend the proposed method to 3D. In addition, advanced tracking techniques can be employed to handle transient deforming structures \cite{zeng2011registration, lee2016landmark}. \blue{Finally, the proposed modal analysis can be adapted to develop reduced-order models for studying different control strategies for morphing systems to achieve better mobility or achieve higher energy extraction efficiency in nonuniform flow conditions. }

\begin{acknowledgments}
This study is partly supported by the Defense Advanced Research Projects Agency, Grant number D19AP00035. The author would also like to thank Florida State University Research Computing Center for providing the computational resources.
\end{acknowledgments}

\appendix

\section{\label{apx:grid}Grid Refinement Study}

\begin{figure*}
	\centering
	\makebox[\textwidth][c]{\includegraphics{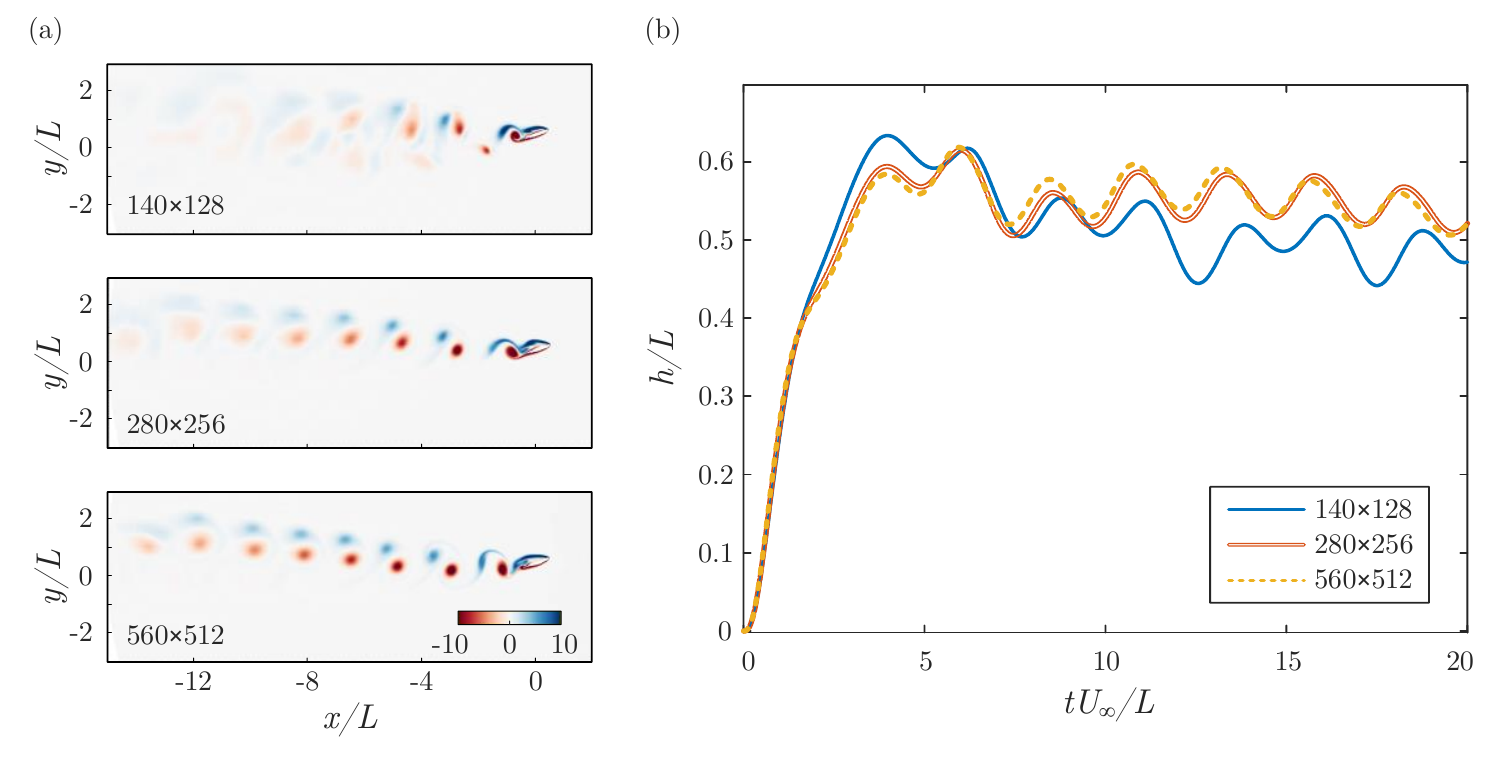}}
	\caption{(a) Instantaneous snapshot of the vorticity field with the resolution of $140\times128$, $280\times256$ and $560\times512$ in the radial and azimuthal directions, respectively. (b) Comparison of the heaving time histories with three different grid resolutions.}
	\label{fig:A1}
\end{figure*}

A grid refinement study is performed by employing coarse ($140\times128$), fine ($560\times512$) and medium ($280\times256$) grid resolutions. Figure \ref{fig:A1}(a) compares the instantaneous snapshots of these resolutions at the same time instance and the same simulation parameters (AoA$=20^\circ$, $\theta_0=22.5^\circ$, $A=22.5^\circ$, $St_f=0.4$). Similar vortex fields are obtained for the medium and fine grids. The heaving displacement time history plotted in Fig. \ref{fig:A1}(b) confirms that the medium resolution can capture the foil movement accurately. Based on this, the medium resolution grid is selected for the current study. We refer to a previous work \cite{wang2021geometrically} for the validation of the algorithm.

\section{\label{apx:stiffness}Effect of Structural Stiffness and Damping}
\begin{figure}
	\centering
	\makebox[\textwidth][c]{\includegraphics[width=1.0\textwidth]{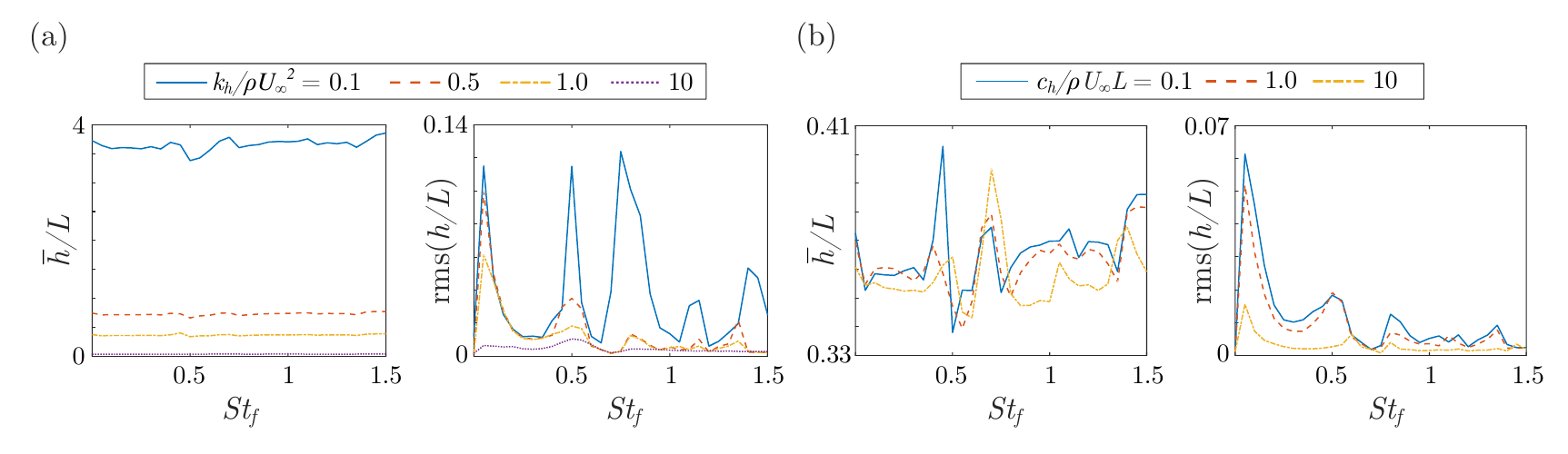}}
	\caption{The average heaving displacement and heaving amplitude of (a) varied stiffness with fixed unity damping and (b) varied damping with fixed unity stiffness.}
	\label{fig:B1}
\end{figure}

Figure \ref{fig:B1} shows the average heaving displacement and oscillatory amplitude for different flap frequencies and varied stiffness and damping values. As anticipated, the average heaving displacement reduces inverse-proportionally with the increase of $k_h$. On the other hand, increasing the damping does not change the average displacement much but only decreases the heaving amplitude. Changing the stiffness or damping does not significantly impact the dominant frequencies, which implies that the flap-induced and the flow-induced vortex shedding are still the main driving force of the plunging motion. Hence, the observation discussed in the current paper could be drawn to a wide range of structural parameters. However, from the preliminary analysis, the structural properties profoundly impact the power generated when the damper is used as the energy harvesting mechanism. The identification of optimal combinations of structural design and active control to achieve the highest energy harvesting efficiency is the subject of ongoing work. 


\bibliography{ref}

\end{document}